\documentclass[letter]{aa} 
\usepackage{graphicx}
\usepackage{txfonts}
\newcommand{\gaia}{\textit{Gaia}\xspace}
\begin{document}

   \title{ Masses of the Hyades white dwarfs }

   \subtitle{A gravitational redshift measurement}

   \author{L. Pasquini 
          \inst{1}
          \and
          A.~F. Pala
          \inst{1} \and
          H.-G. Ludwig \inst{2}\and I.C L\~eao \inst{3} \and J.R. de Medeiros \inst{3} \and Achim Weiss \inst{4} \thanks{Based on UVES data from  the ESO VLT archive}
          }
   \institute{ESO, Karl Schwarzschild Strasse 2, D-85748 Garching, Germany  
       \and Zentrum f\"ur Astronomie der Universit\"at Heidelberg,
   Landessternwarte, K\"onigstuhl 12, D-69117 Heidelberg, Germany
            \and Departamento de F\'isica, Universidade Federal do Rio Grande do Norte, 59078-970 Natal, RN, Brazil  
            \and Max-Planck-Institut f\"ur Astrophysik, Karl-Schwarzschild-Str. 1, D-85748 Garching, Germany }

   \date{}

 
  \abstract
      {
        It is possible to accurately measure the masses of the  white dwarfs (WDs)  in the Hyades cluster using gravitational redshift, because the  radial velocity of the stars can be  obtained independently of spectroscopy from astrometry and the cluster has a low velocity dispersion. }
      {We aim to obtain an accurate   measurement of the  Hyades WD masses
        by determining the mass-to-radius ratio (M/R)\textbf{} from the  observed gravitational redshift, and to compare them with masses derived from other methods. }
      {We analyse archive high-resolution UVES-VLT spectra of six  WDs belonging to the Hyades to measure their Doppler shift, from which M/R  is determined after subtracting the astrometric radial velocity. We estimate the radii using \textit{Gaia} photometry  as well as literature data.   }
   {
     The M/R error associated to the  gravitational redshift measurement is about 5$\%$. 
     The radii estimates, evaluated with different methods, are   in very good agreement, though they can differ by up to 4$\%$ depending on the quality of the data.   The masses based on gravitational redshift are systematically smaller
     than those derived from other methods, by a minimum of $\sim$ 0.02  up to 0.05 solar masses.
     While this difference is  within our measurement uncertainty, the fact that it is systematic indicates a likely real discrepancy between the different methods.   }
   { We show that the M/R derived from gravitational redshift measurements is a powerful tool to  determine the masses of the  Hyades WDs and could reveal interesting properties of their atmospheres.
     The technique can be improved by using dedicated spectrographs, and can be extended to other clusters, making it unique in its ability to
      accurately and empirically determine  the masses of WDs in open clusters. At the same time we prove that gravitational redshift in WDs agrees with the  predictions of stellar evolution models to within a few percent. }

   \keywords{Stars: white dwarfs -- fundamental parameters --
                Techniques: radial velocities 
            }

   \maketitle
%

\section{Introduction}
The importance of the accurate determination of stellar mass cannot be overemphasized. Especially for white dwarfs (WDs), the mass-to-radius relationship is interesting because it is accurately predicted by stellar physics and evolution, and unlike in other stages of stellar evolution, more massive stars are predicted to have smaller radii.
The ratio between the initial stellar mass and the  mass of WDs for instance is essential to model stellar populations, and more generally to verify predictions of stellar-evolution models such as cooling times, ages, and masses of WDs.
A direct comparison between models and observations is therefore fundamental. For example, \citet[see e.g.][]{Salaris+2018} use  evolutionary models to estimate the masses of the Hyades WDs with a precision of about 2$\%$. 
  The radii and masses of WDs can be obtained by observing these targets in eclipsing binaries, to better than  1$\% $ accuracy \citep[see e.g.][]{Parsons+2017}. 
We think that it is now possible to observationally determine the masses of the Hyades WDs with good  accuracy because  
i) \gaia provides very accurate distances (to better than 0.5$\%$) for these stars, and ii)  the Hyades 
cluster radial velocity is known with an uncertainty of  about 100 ms$^{-1}$ \citep{Leao+2019}.  
These latter authors have also shown that spectroscopic and astrometric radial velocities agree to better 
than $\sim$ 30 ms$^{-1}$ once the additional causes of line shifts are properly taken into account: 
gravitational redshift, atmospheric convective motions, and cluster rotation. After applying these corrections,  
the  cluster velocity dispersion is small: less than  340 ms$^{-1}$ \citep{Leao+2019}.  
Therefore, in principle, by measuring the spectral Doppler shift of the WDs in the Hyades, their gravitational redshift (GR) can be 
derived with an error of $\sim$~300 ms$^{-1}$. 
In most stars  the GR is of a few hundred metres per second, comparable to the convective blueshifts in stellar atmospheres \citep[see e.g.][]{Allende+2002, Leao+2019}, but in WDs the GR signal is very pronounced (tens of kms$^{-1}$), and, if hotter than $\sim$ 14\,000 K, no other mechanism capable of shifting the lines of these stars is expected to be present \citep{Allende+2002,Tremblay+2013}.
This implies that the difference between  the spectroscopic Doppler shifts and the astrometric radial velocity can be attributed entirely to  GR, making the measurement potentially very accurate.

White dwarf binaries are also suitable candidates for this technique, provided the binary system is very well characterised, and  GR observations have been used with  success for Sirius B to show the equivalence between the GR mass and dynamical mass of this star \citep{Joyce+2018}. 

The eight well-established DA WDs (white dwarfs with hydrogen atmospheres) belonging to the Hyades cluster \citep{Salaris+2018} are therefore  ideal candidates for GR measurement. 

\section{Doppler shift and GR measurements}

  \begin{figure}
 \centering
\includegraphics[width=0.48\textwidth]{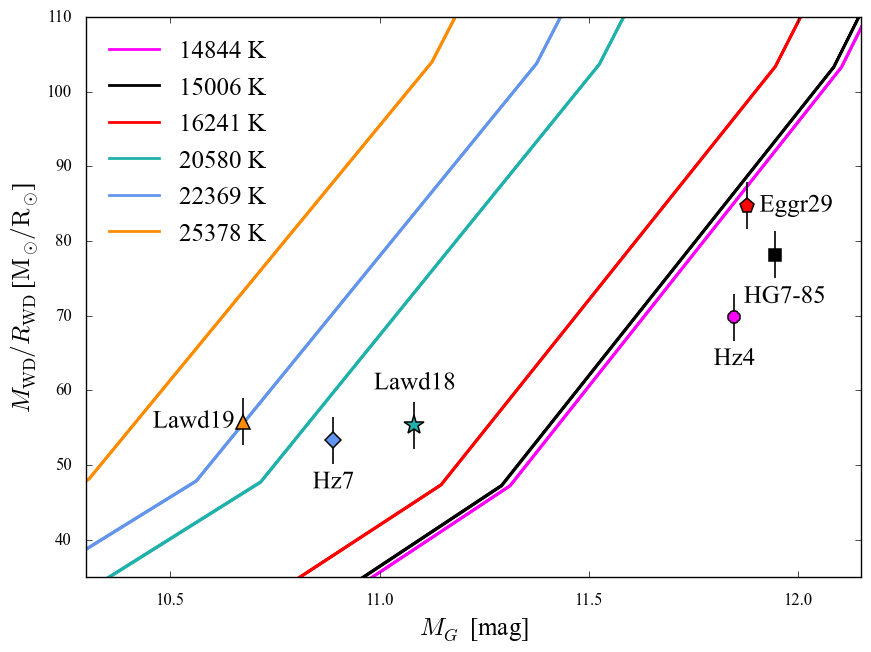} 

   \caption{Comparison between observations and models in an observational plane: M/R vs. \textit{Gaia} $G$ magnitudes. Each star is indicated with a different symbol and colour. For each star, its effective temperature has been computed as the average value between the measurements by \citet{Cummings+2018} and \citet{Gentile+2019} 
   and the corresponding models are over-imposed and colour-coded as the corresponding star. All stars lay below the corresponding curves, indicating a discrepancy between the  adopted models and our observations.}
         \label{Fig1}
  \end{figure}

The spectral lines of WDs are very broad, and in principle this makes the measurement of their Doppler shift  relatively difficult.
However,  the  main Balmer lines (notably H$\alpha$ and sometimes H$\beta$) show non- local thermodynamic equilibrium (LTE) narrow cores, 
allowing precise measurements  of their Doppler shifts. This feature has been used in the past to measure GR or to search for 
binary progenitors of supernovae \citep[see e.g.][]{Reid1996,Napiwotzki+2001,Zuckerman+2013}. 

Out of the eight classical DA WDs in the Hyades, six have been observed with UVES at the 
VLT in the early 2000s by the Supernova-Ia Progenitor Survey (SPY) large programme \citep{Napiwotzki+2001} and they are listed in Table~\ref{tab:wd_gr}.
Each star has been observed at two different epochs.
The signal to noise ratio (S/N) for each observation is about 25 in the continuum close to H$\alpha$. 
All the stars show a  narrow non-LTE H$\alpha$ core, and the combined error arising from the spectra S/N and the line core width  translates into fit measurement errors between 0.8 and 1.5 kms$^{-1}$. The two measurements per star are indeed consistent with these errors. 
The same spectra were previously analysed by \citet{Falcon+2010}, and a comparison between their H$\alpha$ 
Doppler shift measurements and ours shows excellent agreement to within a few hundred metres per second for all stars. 
Since we are interested in absolute measurements we in addition consider that the spectra were 
obtained with a relatively wide slit (2 arcseconds). This adds to the line fit uncertainty the error induced by centering the star in the slit, by a non-predictable amount (note that 1 kms$^{-1}$ corresponds to a shift 
of less than 0.2 arcsecond in the UVES slit, \citealt{Pasquini+2015}). 
Finally, UVES is not in vacuum, nor is it thermally stabilised, and so shifts caused by variations of these environmental parameters are possible.
The reduced data were taken from the UVES science data products in the VLT archive, and the measured Doppler velocities are provided in Table~\ref{tab:wd_gr} (corrected to the solar system barycenter) 
as measured from the H$\alpha$ core and averaging the two observations. 
The associated error  is simply half the difference between the two UVES spectra. 
When adding the other sources of error,  we consider that an overall  uncertainty of  
2 km$^{-1}$ is realistic. We assumed for H$\alpha$ a wavelength of 6562.801 \AA. 

The Doppler shifts were subsequently corrected for the astrometric radial velocity of each star and for the cluster rotation, though the latter is less than 300 ms$^{-1}$ for the sample stars \citep{Leao+2019}, providing the GR for each star. All the values obtained are given in Table~\ref{tab:wd_gr} . 

In the same table we also provide the Doppler velocities  measured by \citet{Zuckerman+2013} and the GR measured by \citet{Reid1996} with HIRES at Keck.
These independent measurements agree well with ours, 
confirming that the assumed 2 kms$^{-1}$ uncertainty is very reasonable. 
It is worth noting that  \citet{Zuckerman+2013} used the blue setup of HIRES, which
does not provide access to H$\alpha$, and higher lines in the Balmer series were measured. 
These latter lines do not  show narrow cores, explaining why their velocities have a comparable uncertainty 
to the UVES ones, in spite of the fact that the Keck spectra  are at a higher resolution and have higher S/N ratios. 
Since other lines  as well as another spectrograph were used, the \citeauthor{Zuckerman+2013}  
measurements are  complementary and independent from ours. The maximum difference between the \citeauthor{Zuckerman+2013} Doppler shift and ours is of 3 kms$^{-1}$, in agreement with our estimated uncertainty. 

We finally compute M/R (in solar unit) by averaging the UVES and \citeauthor{Zuckerman+2013} shifts (also reported in Table~\ref{tab:wd_gr}) and using the formula: GR=0.6365(M/R) (in kms$^{-1}$). The M/R values are reported in the last column of Table~\ref{tab:wd_gr}.
 We  expect an error of about 3 units in  M/R, which translates into a 4-5$\%$ uncertainty. 

 \begin{table*}
      \caption[]{Gravitational redshift measurement. Column 1: star id; column 2: measured Doppler shift from UVES spectra, averaged between  two observations; column 3: 
      half difference between the two UVES observations; column 4: 
      Doppler shift as measured by \citet{Zuckerman+2013} from Keck spectra; column 5: UVES and \citeauthor{Zuckerman+2013} 
      average Doppler shifts; column 6: astrometric radial velocities; column 7: correction for cluster rotation; column 8: GR (from column 5,6 and 7); column 9: GR from \citet{Reid1996} Keck spectra; column 10: M/R from our GR estimate. All columns are expressed in kms$^{-1}$.  }\label{tab:wd_gr}
         \label{redshift}
         \centering          
\begin{tabular}{c c c c c c c c c c }     
\hline\hline       
            Star  & Doppler Shift & $\Delta$v & Zucker & Ave   & RV(astro) & Rotation & GR  & Reid GR & M/R  \\
            HZ4   &   82.64   & 0.7   & 79.4  & 81.0  & 36.29  & -0.27   &  44.4   & 41.6 & 69.76 \\
            HZ7    &  75.07   & 1.0   & 73.3  & 74.2  & 40.39  &  0.06   &  33.9   & 33.5 & 53.26   \\
            LAWD19 &  74.52   & 0.2   & 75.4  & 75.0  & 39.55  &  0.01   &  35.5   & 35.4 & 55.77 \\
            LAWD18  & 73.87   & 0.8   & 75.0  & 74.4  & 39.15  & -0.03   &  35.2   & 35.7 & 55.30  \\
            EGGR29  & 90.16   & 1.0   & 93.1  & 91.6  & 37.54  & -0.15   &  53.9   & 52.3 & 84.68  \\
            HG7-85  & 87.81   & 2.0   & 86.3  & 87.0 & 37.07 & -0.20     &  49.7   &      & 78.08  \\

     \hline

    \end{tabular}
   \end{table*}

 \section{The estimate of the radii}

 Once M/R has been measured, we need to estimate the stellar radii.
 Since the angular extension of the stars is too small for interferometric measurements,  
 the most direct way of determining the stellar radii is to compute bolometric magnitudes and apply the definition of 
 luminosity.
 Temperatures and stellar parameters can also be obtained  by the  spectral fitting of the Balmer lines and the global fitting of the spectra 
 (hereafter `spectroscopic', e.g. \citealt{Tremblay+2012,Gianninas+2011}).
 Following \citet{Salaris+2018}, we  used \textit{Gaia}
 colours and magnitudes, and the WD  model atmospheres  developed by Bergeron and collaborators 
  \citep{Holberg2006, Kowalski2006, Tremblay+2011} by interpolating the models for DA dwarfs and thin hydrogen atmospheres.
 Using the \gaia parallaxes, we computed the absolute magnitudes in the \gaia filter, 
 $G$, $G_\mathrm{BP}$ and $G_\mathrm{RP}$, and performed a fit to the 
 synthetic magnitudes kindly provided by Dr. Bergeron. The best-fit model provides a 
 measurement of the effective temperature ($T_\mathrm{eff}$) and the bolometric magnitude ($M_\mathrm{bol}$) 
 of each WD. Given the 
 vicinity of the Hyades, we did not apply any reddening correction.
%
%
Assuming the standard values for the Sun, $T_\odot = 5780\,$K and $M_\mathrm{bol \odot} = 4.75\,$mag, the radius of the WD can then be determined by combining 
the bolometric magnitude definition with the Stefan-Boltzmann law. The results are given in Table~\ref{radii}.

 Radii for these stars have been estimated, either from spectroscopy 
 (e.g. \citealt{Cummings+2018}, whose results are very similar to those of \citealt{Tremblay+2012}), or 
 from photometry, from either \textit{Gaia} or the Sloan Digital Sky Survey (SDSS)  \citep{Gentile+2019}, 
 and they are also reported in Table~\ref{radii}. 
 The agreement between the radii estimates  is good, but they may differ by up to  10 $\%$  (peak to peak) for some of the stars. The  two radii based on SDSS photometry by \citet{Gentile+2019} are systematically smaller than the other values,  but we have no explanation for this (see also \citealt{Bergeron+2019}).  We have not used these values in our radii estimates.
 
For  HZ4 and EGGR29 the agreement is excellent, independent of the diagnostic used. Since  these two stars are spectrophotometric  standards, we expect the quality of their photometry and spectroscopy to be superior. The excellent agreement  indicates  that the discrepancies likely reflect the quality of the input data. 

Following the method described above, we also computed the WD radii, performing the photometric fit to the SED composed by the Johnson $V$ magnitudes and $B-V$ colours, the \gaia photometry, and the SDSS photometry,  whenever available. The results are listed in column $R_\mathrm{j+s+g}$ of Table~\ref{radii}.
For two stars this would produce larger radii by about 0.006 R$_\mathrm{\odot}$, while for the others the agreement with our average is excellent. 

 Gravitational redshift masses are  given in Table~\ref{radii}, and they have been computed using the M/R ratios determined in the previous section and the radii computed by averaging the \gaia values from \citet{Gentile+2019} (column 2), \citeauthor{Cummings+2018} (column 6), and our \gaia values (column 8).  We have not used the SED estimates (column 9) in making the average because they include the \gaia colours.
 The average is reported in Table~\ref{radii} (column 10), together with the associated  dispersion. We  notice that this dispersion is not a realistic estimate of the uncertainty; the reason being that the radii estimates are  not independent, since two are based on the same \gaia colours and all use the same family of model-based transformations. In addition, the results from \citet{Gentile+2019} and \citet{Cummings+2018} implicitly use  the mass-to-radius relationship of \citet{Fontaine+2001}.  

 \begin{table*}
   \caption[]{Stellar radii and masses. Column 2:  radius from \citet{Gentile+2019}  \textit{Gaia}-based gravity and mass; column 3: masses as reported by \citet{Gentile+2019} for the \textit{Gaia} analysis;
     columns 4 and 5: radii and masses from the same authors, based on SDSS colour analysis;
     columns 6 and 7: radii and masses as from \citet{Cummings+2018}; columns 8 and 9: radii derived by us on \gaia photometry and on SED fitting
     ; column 10: average radii from  columns  2,6 and 8; column 11: masses derived from average radii of column 10 and GR from Table~\ref{tab:wd_gr}.
     Labels a, b, ad c in column 9 refer to different sources of Johnson photometry: a \citet{Landolt2007}, b \citet{Tremblay+2012}, c  \citet{Zacharias+2012}. }
         \label{radii}
          \centering          
          \begin{tabular}{c c c c c c c c c c  c}     
            \hline\hline                               
            Star       &   RGF$_g$ & MGF$_g$ & RGF$_{s}$ & MGF$_s$ & RCum   & MCum    & R$_g$  & R$_{j+s+g}$  & Radius    & Mass $_{gr}$ \\
\hline
            HZ4        &   0.0104   & 0.807(6) &         &         & 0.0105 &  0.797  &  0.0105   & 0.0107 a  & 0.0105(1) &  0.73(4) \\
            HZ7        &   0.0119   & 0.713(9) &         &         & 0.0121 &  0.691  &  0.0124   & 0.0126 b  & 0.0121(2) &  0.64(6) \\
            LAWD19     &   0.0116   & 0.74(1)  &         &         & 0.0121 &  0.704  &  0.0120   & 0.0126 b  & 0.0119(2) &  0.66(5) \\ 
            LAWD18     &   0.0115   & 0.735(8) & 0.0099  & 0.82    & 0.0119 &  0.700  &  0.0119   & 0.0117 b  & 0.0118(1) &  0.65(5) \\
            EGGR29     &   0.0098   & 0.86(1)  &         &         & 0.0099 &  0.850  &  0.0101   & 0.0100 a  & 0.0099(1) &  0.84(3) \\
            HG7-85     &   0.0099   & 0.845(7) & 0.0097  & 0.836   & 0.0109 &  0.765  &  0.0101   & 0.0103 c  & 0.0103(3) &  0.80(4) \\
           
            \hline
   \end{tabular}
   \end{table*}

 In the determination of the GR mass, we  estimate an uncertainty of about 0.03 M$_{\odot}$ introduced by the error on the Doppler shift. The uncertainty on the radius is more difficult to  evaluate, 
 and the values in Table~\ref{radii}, obtained with  different methods and colours, differ from  a minimum of $\sim$ 1 $\%$
 for the most well-characterised stars to 4 $\%$ for the most poorly characterised, not considering the SDSS-based estimate of \citet{Gentile+2019}. 
 
 Bearing these uncertainties  in mind,  our GR  masses are systematically smaller than those estimated by other authors.  
 
 A similar systematic shift is present for five stars 
 when comparing our GR masses with those derived by \citet{Salaris+2018}, who claim an accuracy of $\sim$ 0.01 M$_{\odot}$.
 
 The uncertainty associated to our GR determination is too large to derive a firm conclusion, but
 it is unlikely that  all our velocities are systematically too small. Furthermore, there is no  evidence of a systematic effect when 
 comparing our velocities with those obtained  by other authors (see Table~\ref{tab:wd_gr}).
 Could the WDs in the Hyades have different kinematics from the rest of the Hyades stars?  
 In their study of the WDs in the globular cluster NGC6397, \citet{davis+2008} argue that asymmetric kicks could explain the peculiar kinematics of these stars, 
 but  these kicks should result in a larger velocity dispersion for this group of stars, not in a systematic effect. 
 Effects, such as mass segregation are also expected to enhance the velocity dispersion rather than produce asymmetries, and \citet{Leao+2019}  
 find  that the cluster radial velocity remains substantially unchanged when sampling with different radii. 

 Furthering our case in Figure~\ref{Fig1} we show two observational quantities, namely M/R as measured from GR and\ the absolute stellar $G$ magnitude, with the associated  error bars. For each star we computed the effective temperature averaging from the \citep{Gentile+2019} and the \citep{Cummings+2018} values, and the tracks corresponding  to these temperatures are also displayed, with the same colour as the star symbol.   Clearly  the observed points would agree with models systematically cooler than what was found. 
 We can therefore safely conclude that our GR masses, although not of adequate accuracy, are lower than what has been found by  other studies.
 While we can quantify the error in the Doppler shifts, the one in the radius  is more  difficult to estimate, 
 because we have no control on the systematic effects in the models. The use of different techniques and data seem 
 to produce differences  of up to 0.001 R$_\mathrm{\odot}$, or 10 $\%$  (peak to peak) in the worse case.  
 
\citet{romero+2019} study the dependence of the WD parameters from the stellar atmospheres, finding that the radii 
of WDs decrease the smaller the content of hydrogen in their envelopes,   by up to several percent for stars with masses 
and temperatures comparable to the Hyades WDs.  If the mass of the H layer was smaller than the 10$^{-4}$ M$_{\odot}$ 
adopted by both \citet{Cummings+2018} and \citet{Gentile+2019}, then their gravity would be systematically overestimated 
and consequently the masses derived would also be higher. It seems therefore quite possible that the discrepancy between 
these authors and our GR-based measurement is due to a lower-mass H envelope, but we feel that given the associated 
uncertainties,  pushing such a hypothesis further is beyond the scope of this work. 
 
 \section{Conclusions}

 The classical DA WDs in the Hyades cluster  are hot enough to avoid complications from atmospheric effects (convective shifts). Moreover, since the cluster velocity is determined with high accuracy, 
and  the equivalence between astrometric and spectroscopic radial velocities has been shown to better than 30 ms$^{_1}$ for the Hyades stars,  astrometric velocities can be directly subtracted from the measured Doppler shifts{Pleas. 
 This,  coupled with the low internal velocity  dispersion of the cluster, make them ideal candidates for measurement of their masses with GR.

 We used archive VLT-UVES observations to measure the GR of six bona-fide Hyades WDs, showing that even if these observations were  taken for other purposes,
 they can  define the M/R ratio to within an accuracy of 5$\%$. 

 We estimated the stellar radii using 
 photometry and models, and we also used literature results, either based on photometry or on spectroscopy. The radii estimates can vary, 
 and we find that the two SDSS-based radii \citep{Gentile+2019} tend to give smaller radii than the others.
 The agreement between the different estimates varies between 1$\%$ for the best cases and  3$\%$ (rms) for the worst. 

 The masses 
 based on  gravitational redshifts are systematically smaller than those found in recent literature. 
 A small systematic difference exists also  with the  values estimated by \citet{Salaris+2018} based on stellar evolution models.

The GR measurement can be improved by using dedicated instruments and observations, such as the recently commissioned ESPRESSO at the VLT \citep{Pepe+2010}, which provides enormous 
advantages thanks to several technical solutions (stability of input, the atmospheric dispersion corrector, and the vacuum instrument). 
Nevertheless, our results show how powerful the method can be
when applied to suitable open clusters. 
While for other  open clusters the situation will be less favourable than for the Hyades, 
still the uncertainty in their systemic spectroscopic velocities can be of less than 1 kms$^{_1}$, 
which induces a typical uncertainty of less than 2$\%$ on mass estimates for WDs.  This compares very well 
with alternative methods and it has the potential to approach the accuracy obtained with eclipsing 
binaries \citep{Parsons+2017}.

We finally  note  that, assuming we know the mass of the WDs from evolutionary and spectroscopic studies, 
as published in literature, our GR measurements lead to mass values that show agreement to within a few percent with evolutionary and gravitational masses, 
certainly within the measurement errors. We have therefore tested GR in the gravity regime of  WDs to an accuracy of a few percent. 
There are not many stellar GR measurements present in the literature for WDs, and usually the precision of the measured Doppler shifts is limited. 
Thus  we have brought the consistency between GR and stellar evolution  into a novel domain \citep{Parsons+2017},  
that has thus far only been explored for Sirius B  \citep{Joyce+2018}.

\begin{acknowledgements}
  HGL acknowledges financial support by the Sonderforschungsbereich SFB\,881
``The Milky Way System'' (subprojects A4) of the German Research Foundation
(DFG). Research activities of the Observational Astronomy Board of the Federal University of Rio Grande do Norte (UFRN) are supported by continuing grants from the Brazilian agencies CNPq, CAPES and FAPERN. We acknowledge the

 anonymous referee for several suggestions, especially for pointing out the influence of the mass of the H envelope. 
\end{acknowledgements}

%
\bibliographystyle{aa} 
\bibliography{wd2} 

\begin{thebibliography}{25}
\expandafter\ifx\csname natexlab\endcsname\relax\def\natexlab#1{#1}\fi

\bibitem[{{Allende Prieto} {et~al.}(2002){Allende Prieto}, {Lambert}, {Tull},
  \& {MacQueen}}]{Allende+2002}
{Allende Prieto}, C., {Lambert}, D.~L., {Tull}, R.~G., \& {MacQueen}, P.~J.
  2002, \apjl, 566, L93

\bibitem[{{Bergeron} {et~al.}(2019){Bergeron}, {Dufour}, {Fontaine}, {Coutu},
  {Blouin}, {Genest-Beaulieu}, {B{\'e}dard}, \& {Rolland}}]{Bergeron+2019}
{Bergeron}, P., {Dufour}, P., {Fontaine}, G., {et~al.} 2019, arXiv e-prints
  [\eprint[arXiv]{1904.02022}]

\bibitem[{{Cummings} {et~al.}(2018){Cummings}, {Kalirai}, {Tremblay},
  {Ramirez-Ruiz}, \& {Choi}}]{Cummings+2018}
{Cummings}, J.~D., {Kalirai}, J.~S., {Tremblay}, P.-E., {Ramirez-Ruiz}, E., \&
  {Choi}, J. 2018, \apj, 866, 21

\bibitem[{{Davis} {et~al.}(2008){Davis}, {Richer}, {King}, {Anderson},
  {Coffey}, {Fahlman}, {Hurley}, \& {Kalirai}}]{davis+2008}
{Davis}, D.~S., {Richer}, H.~B., {King}, I.~R., {et~al.} 2008, \mnras, 383, L20

\bibitem[{{Falcon} {et~al.}(2010){Falcon}, {Winget}, {Montgomery}, \&
  {Williams}}]{Falcon+2010}
{Falcon}, R.~E., {Winget}, D.~E., {Montgomery}, M.~H., \& {Williams}, K.~A.
  2010, \apj, 712, 585

\bibitem[{{Fontaine} {et~al.}(2001){Fontaine}, {Brassard}, \&
  {Bergeron}}]{Fontaine+2001}
{Fontaine}, G., {Brassard}, P., \& {Bergeron}, P. 2001, \pasp, 113, 409

\bibitem[{{Gentile Fusillo} {et~al.}(2019){Gentile Fusillo}, {Tremblay},
  {G{\"a}nsicke}, {Manser}, {Cunningham}, {Cukanovaite}, {Hollands}, {Marsh},
  {Raddi}, {Jordan}, {Toonen}, {Geier}, {Barstow}, \&
  {Cummings}}]{Gentile+2019}
{Gentile Fusillo}, N.~P., {Tremblay}, P.-E., {G{\"a}nsicke}, B.~T., {et~al.}
  2019, \mnras, 482, 4570

\bibitem[{{Gianninas} {et~al.}(2011){Gianninas}, {Bergeron}, \&
  {Ruiz}}]{Gianninas+2011}
{Gianninas}, A., {Bergeron}, P., \& {Ruiz}, M.~T. 2011, \apj, 743, 138

\bibitem[{{Holberg} \& {Bergeron}(2006)}]{Holberg2006}
{Holberg}, J.~B. \& {Bergeron}, P. 2006, \aj, 132, 1221

\bibitem[{{Joyce} {et~al.}(2018){Joyce}, {Barstow}, {Holberg}, {Bond},
  {Casewell}, \& {Burleigh}}]{Joyce+2018}
{Joyce}, S.~R.~G., {Barstow}, M.~A., {Holberg}, J.~B., {et~al.} 2018, \mnras,
  481, 2361

\bibitem[{{Kowalski} \& {Saumon}(2006)}]{Kowalski2006}
{Kowalski}, P.~M. \& {Saumon}, D. 2006, \apjl, 651, L137

\bibitem[{{Landolt} \& {Uomoto}(2007)}]{Landolt2007}
{Landolt}, A.~U. \& {Uomoto}, A.~K. 2007, \aj, 133, 768

\bibitem[{{Le{\~a}o} {et~al.}(2019){Le{\~a}o}, {Pasquini}, {Ludwig}, \& {de
  Medeiros}}]{Leao+2019}
{Le{\~a}o}, I.~C., {Pasquini}, L., {Ludwig}, H.-G., \& {de Medeiros}, J.~R.
  2019, \mnras, 483, 5026

\bibitem[{{Napiwotzki} {et~al.}(2001){Napiwotzki}, {Christlieb}, {Drechsel},
  {Hagen}, {Heber}, {Homeier}, {Karl}, {Koester}, {Leibundgut}, {Marsh},
  {Moehler}, {Nelemans}, {Pauli}, {Reimers}, {Renzini}, \&
  {Yungelson}}]{Napiwotzki+2001}
{Napiwotzki}, R., {Christlieb}, N., {Drechsel}, H., {et~al.} 2001,
  Astronomische Nachrichten, 322, 411

\bibitem[{{Parsons} {et~al.}(2017){Parsons}, {G{\"a}nsicke}, {Marsh}, {Ashley},
  {Bours}, {Breedt}, {Burleigh}, {Copperwheat}, {Dhillon}, {Green}, {Hardy},
  {Hermes}, {Irawati}, {Kerry}, {Littlefair}, {McAllister}, {Rattanasoon},
  {Rebassa-Mansergas}, {Sahman}, \& {Schreiber}}]{Parsons+2017}
{Parsons}, S.~G., {G{\"a}nsicke}, B.~T., {Marsh}, T.~R., {et~al.} 2017, \mnras,
  470, 4473

\bibitem[{{Pasquini} {et~al.}(2015){Pasquini}, {Cort{\'e}s}, {Lombardi},
  {Monaco}, {Le{\~a}o}, \& {Delabre}}]{Pasquini+2015}
{Pasquini}, L., {Cort{\'e}s}, C., {Lombardi}, M., {et~al.} 2015, \aap, 574, A76

\bibitem[{{Pepe} {et~al.}(2010){Pepe}, {Cristiani}, {Rebolo Lopez}, {Santos},
  {Amorim}, {Avila}, {Benz}, {Bonifacio}, {Cabral}, {Carvas}, {Cirami},
  {Coelho}, {Comari}, {Coretti}, {De Caprio}, {Dekker}, {Delabre}, {Di
  Marcantonio}, {D'Odorico}, {Fleury}, {Garc{\'{\i}}a}, {Herreros Linares},
  {Hughes}, {Iwert}, {Lima}, {Lizon}, {Lo Curto}, {Lovis}, {Manescau},
  {Martins}, {M{\'e}gevand}, {Moitinho}, {Molaro}, {Monteiro}, {Monteiro},
  {Pasquini}, {Mordasini}, {Queloz}, {Rasilla}, {Rebord{\~a}o}, {Santana
  Tschudi}, {Santin}, {Sosnowska}, {Span{\`o}}, {Tenegi}, {Udry}, {Vanzella},
  {Viel}, {Zapatero Osorio}, \& {Zerbi}}]{Pepe+2010}
{Pepe}, F.~A., {Cristiani}, S., {Rebolo Lopez}, R., {et~al.} 2010, in
  \procspie, Vol. 7735, Ground-based and Airborne Instrumentation for Astronomy
  III, 77350F

\bibitem[{{Reid}(1996)}]{Reid1996}
{Reid}, I.~N. 1996, \aj, 111, 2000

\bibitem[{{Romero} {et~al.}(2019){Romero}, {Kepler}, {Joyce}, {Lauffer}, \&
  {C{\'o}rsico}}]{romero+2019}
{Romero}, A.~D., {Kepler}, S.~O., {Joyce}, S.~R.~G., {Lauffer}, G.~R., \&
  {C{\'o}rsico}, A.~H. 2019, \mnras, 484, 2711

\bibitem[{{Salaris} \& {Bedin}(2018)}]{Salaris+2018}
{Salaris}, M. \& {Bedin}, L.~R. 2018, \mnras, 480, 3170

\bibitem[{{Tremblay} {et~al.}(2011){Tremblay}, {Bergeron}, \&
  {Gianninas}}]{Tremblay+2011}
{Tremblay}, P.-E., {Bergeron}, P., \& {Gianninas}, A. 2011, \apj, 730, 128

\bibitem[{{Tremblay} {et~al.}(2013){Tremblay}, {Ludwig}, {Steffen}, \&
  {Freytag}}]{Tremblay+2013}
{Tremblay}, P.-E., {Ludwig}, H.-G., {Steffen}, M., \& {Freytag}, B. 2013, \aap,
  559, A104

\bibitem[{{Tremblay} {et~al.}(2012){Tremblay}, {Schilbach}, {R{\"o}ser},
  {Jordan}, {Ludwig}, \& {Goldman}}]{Tremblay+2012}
{Tremblay}, P.-E., {Schilbach}, E., {R{\"o}ser}, S., {et~al.} 2012, \aap, 547,
  A99

\bibitem[{{Zacharias} {et~al.}(2012){Zacharias}, {Finch}, {Girard}, {Henden},
  {Bartlett}, {Monet}, \& {Zacharias}}]{Zacharias+2012}
{Zacharias}, N., {Finch}, C.~T., {Girard}, T.~M., {et~al.} 2012, VizieR Online
  Data Catalog, 1322

\bibitem[{{Zuckerman} {et~al.}(2013){Zuckerman}, {Xu}, {Klein}, \&
  {Jura}}]{Zuckerman+2013}
{Zuckerman}, B., {Xu}, S., {Klein}, B., \& {Jura}, M. 2013, \apj, 770, 140

\end{thebibliography}


\begin{thebibliography}{5}
\expandafter\ifx\csname natexlab\endcsname\relax\def\natexlab#1{#1}\fi

\bibitem[{{Le{\~a}o} {et~al.}(2019){Le{\~a}o}, {Pasquini}, {Ludwig}, \& {de
  Medeiros}}]{Leao+2019}
{Le{\~a}o}, I.~C., {Pasquini}, L., {Ludwig}, H.-G., \& {de Medeiros}, J.~R.
  2019, \mnras, 483, 5026

\bibitem[{{Napiwotzki} {et~al.}(2001){Napiwotzki}, {Christlieb}, {Drechsel},
  {Hagen}, {Heber}, {Homeier}, {Karl}, {Koester}, {Leibundgut}, {Marsh},
  {Moehler}, {Nelemans}, {Pauli}, {Reimers}, {Renzini}, \&
  {Yungelson}}]{Napiwotzki+2001}
{Napiwotzki}, R., {Christlieb}, N., {Drechsel}, H., {et~al.} 2001,
  Astronomische Nachrichten, 322, 411

\bibitem[{{Reid}(1996)}]{Reid1996}
{Reid}, I.~N. 1996, \aj, 111, 2000

\bibitem[{{Salaris} \& {Bedin}(2018)}]{Salaris+2018}
{Salaris}, M. \& {Bedin}, L.~R. 2018, \mnras, 480, 3170

\bibitem[{{Zuckerman} {et~al.}(2013){Zuckerman}, {Xu}, {Klein}, \&
  {Jura}}]{Zuckerman+2013}
{Zuckerman}, B., {Xu}, S., {Klein}, B., \& {Jura}, M. 2013, \apj, 770, 140

\end{thebibliography}
%

\end{document}